\documentclass[12pt]{iopart}
\usepackage{tabulary,graphicx}
\usepackage[perpage]{footmisc}
    
\makeatletter
\expandafter\let\csname equation*\endcsname\relax
\expandafter\let\csname endequation*\endcsname\relax
\usepackage{amsmath}
\expandafter\let\csname subarray\endcsname\relax
\expandafter\let\csname endsubarray\endcsname\relax
\expandafter\let\csname substack\endcsname\relax
\expandafter\let\csname endsubstack\endcsname\relax
\def\[{\relax\ifmmode\@badmath\else
 \begin{trivlist}
 \@beginparpenalty\predisplaypenalty
 \@endparpenalty\postdisplaypenalty
 \item[]\leavevmode
 \hbox to\linewidth\bgroup$ \displaystyle
 \hskip\mathindent\bgroup\fi}
\def\]{\relax\ifmmode \egroup $\hfil \egroup \end{trivlist}\else \@badmath \fi}
\def\equation{\@beginparpenalty\predisplaypenalty
 \@endparpenalty\postdisplaypenalty
\refstepcounter{equation}\trivlist \item[]\leavevmode
 \hbox to\linewidth\bgroup $ \displaystyle
\hskip\mathindent}
\def\endequation{$\hfil \displaywidth\linewidth\@eqnnum\egroup \endtrivlist}
\@namedef{equation*}{\[}
\@namedef{endequation*}{\]}
\def\eqnarray{\stepcounter{equation}\let\@currentlabel=\theequation
\global\@eqnswtrue
\global\@eqcnt\z@\tabskip\mathindent\let\\=\@eqncr
\abovedisplayskip\topsep\ifvmode\advance\abovedisplayskip\partopsep\fi
\belowdisplayskip\abovedisplayskip
\belowdisplayshortskip\abovedisplayskip
\abovedisplayshortskip\abovedisplayskip
$$\halign to
\linewidth\bgroup\@eqnsel$\displaystyle\tabskip\z@
 {##{}}$&\global\@eqcnt\@ne $\displaystyle{{}##{}}$\hfil
 &\global\@eqcnt\tw@ $\displaystyle{{}##}$\hfil
 \tabskip\@centering&\llap{##}\tabskip\z@\cr}
\def\endeqnarray{\@@eqncr\egroup
 \global\advance\c@equation\m@ne$$\global\@ignoretrue }
		
\usepackage{iopams,setstack}

\usepackage{url,multirow,morefloats,floatflt,cancel,textcomp,tfrupee}
\usepackage{pifont}
\usepackage{float}
\usepackage[latin1]{inputenc}
\usepackage[nointegrals]{wasysym}
\makeatletter

\AtBeginDocument{
\expandafter\ifx\csname eqalign\endcsname\relax
\def\eqalign#1{\null\vcenter{\def\\{\cr}\openup\jot\m@th
  \ialign{\strut$\displaystyle{##}$\hfil&$\displaystyle{{}##}$\hfil
      \crcr#1\crcr}}\,}
\fi
}

\let\lt=<
\let\gt=>
\def\processVert{\ifmmode|\else\textbar\fi}

\@ifundefined{subparagraph}{
\def\subparagraph{\@startsection{paragraph}{5}{2\parindent}{0ex plus 0.1ex minus 0.1ex}%
{0ex}{\normalfont\small\itshape}}%
}{}

\newcommand\role[1]{\unskip}
\newcommand\aucollab[1]{\unskip}
  
\@ifundefined{tsGraphicsScaleX}{\gdef\tsGraphicsScaleX{1}}{}
\@ifundefined{tsGraphicsScaleY}{\gdef\tsGraphicsScaleY{.9}}{}
\def\checkGraphicsWidth{\ifdim\Gin@nat@width>\textwidth
	\tsGraphicsScaleX\textwidth\else\Gin@nat@width\fi}

\def\checkGraphicsHeight{\ifdim\Gin@nat@height>.9\textheight
	\tsGraphicsScaleY\textheight\else\Gin@nat@height\fi}

\def\fixFloatSize#1{\@ifundefined{processdelayedfloats}{\setbox0=\hbox{\includegraphics{#1}}\ifnum\wd0<\columnwidth\relax\renewenvironment{figure*}{\begin{figure}}{\end{figure}}\fi}{}}
\let\ts@includegraphics\includegraphics

\def\inlinegraphic[#1]#2{{\edef\@tempa{#1}\edef\baseline@shift{\ifx\@tempa\@empty0\else#1\fi}\edef\tempZ{\the\numexpr(\numexpr(\baseline@shift*\f@size/100))}\protect\raisebox{\tempZ pt}{\ts@includegraphics{#2}}}}

\AtBeginDocument{\def\includegraphics{\@ifnextchar[{\ts@includegraphics}{\ts@includegraphics[width=\checkGraphicsWidth,height=\checkGraphicsHeight,keepaspectratio]}}}

\def\URL#1#2{\@ifundefined{href}{#2}{\href{#1}{#2}}}

\def\UrlOrds{\do\*\do\-\do\~\do\'\do\"\do\-}%
\g@addto@macro{\UrlBreaks}{\UrlOrds}
\makeatother

\begin{document}

\title[Designing Quantum Router in IBM Quantum Computer]{Designing Quantum Router in IBM Quantum Computer}

\author{Bikash K. Behera$^{1}$,
                Tasnum Reza$^{1}$,
                Angad Gupta$^{1}$ and Prasanta K. Panigrahi$^{1}$}
\address{$^{1}$Department of Physical Sciences\unskip, Indian Institute of Science Education and Research Kolkata\unskip, Mohanpur\unskip, 741246\unskip, West Bengal\unskip, India}

\ead{pprasanta@iiserkol.ac.in}

\begin{abstract}
Quantum router is an essential ingredient in a quantum network. Here, we propose a new quantum circuit for designing quantum router by using IBM's five-qubit quantum computer. We design an equivalent quantum circuit, by the means of single-qubit and two-qubit quantum gates, which can perform the operation of a quantum router. Here, we show the routing of signal information in two different paths (two signal qubits) which is directed by a control qubit. According to the process of routing, the signal information is found to be in a coherent superposition of two paths. We demonstrate the quantum nature of  the router by illustrating the entanglement between the control qubit and the two signal qubits (two paths), and confirm the well preservation of the signal information in either of the two paths after the routing process. We perform quantum state tomography to verify the generation of entanglement and preservation of information. It is found that the experimental results are obtained with good fidelity.  
\end{abstract}

\vspace{2pc}\noindent\textit{Keywords: }{IBM Quantum Experience, Quantum Router, Quantum Communication}
    
\section{Introduction}
Quantum communication is the process of transferring quantum states from one place to another. It plays an important role in the field of quantum information processing \unskip~\cite{122770:4133042}. Communication networks are the indispensable technology to transmit quantum information over long distances among the parties connected in a network. Although, classical laws of physics are used for classical communication, it has been predicted that applying the principles of quantum physics and quantum information can enhance the efficiency of communication devices \unskip~\cite{122770:4133066,122770:4133067,122770:4133068} significantly even if using the similar resources and network architecture \unskip~\cite{122770:4131129,122770:4131173}. Quantum internet has been proposed by Kimble \unskip~\cite{122770:4132995}, which shows a significant improvement in the area of quantum communication from both the theoretical and experimental aspects. The most notable result has been observed in quantum cryptography \unskip~\cite{122770:4113608,122770:4113609}, which can be used for unconditional secure transmission of information. Quantum effects such as entanglement \unskip~\cite{122770:4113811} and the probabilistic nature of measurement are the key mechanisms in achieving a secure quantum communication network.   

Correct routing of signal from its source to the destination is necessary in a complex network architecture for both classical and quantum communication \unskip~\cite{122770:4113768,122770:4113810}. Classical routers allow transmission of signal information which is directed by control information in a classical network \unskip~\cite{122770:4114158}. It is known from classical networks that the impossibility of perfect cloning prevents multi-directional broadcast in a quantum network. Hence, quantum routing needs more elaborate protocols as in contrast to classical routing, any arbitrary quantum information can not be perfectly cloned \unskip~\cite{122770:4114161}. However, theoretically and experimentally approximate cloning has been studied extensively \unskip~\cite{122770:4114286,122770:4114328}, resulting in establishing and implementing optimal cloning protocols for a wide class of qubit distributions \unskip~\cite{122770:4114329,122770:4114330}.    

Quantum network has a wide range of applications \unskip~\cite{122770:4114587}. A router being the key element in a network, uses a control bit to decide the path of transmission for the the signal bit. In a quantum router, both the signal and control bits are represented as the quantum bits which is in general stored in a superposition state, and the control qubit has the potential to control the path of the signal qubit in a coherent superposition of multiple paths, which provides remarkable opportunity as compared to its classical counterpart \unskip~\cite{122770:4129601,122770:4129602}. The quantum routing process enables to realize key applications such as quantum random access memory \unskip~\cite{122770:4129685}, quantum machine learning \unskip~\cite{122770:4129686,122770:4129687} that uses a large sets of data. 

A genuine quantum router satisfies the following six requirements. 

(1) Both the control and signal information are stored in quantum bits. 

(2) The signal information remains unchanged under the routing process. 

(3) The router has to be able to route the signal in a coherent superposition of both the output modes. 

(4) The router has to work without any need for post-selection of signal qubits. 

(5) To optimize the resources of quantum network, only one control qubit is required to direct the signals. 

(6) To demonstrate the quantum nature of the quantum router, entanglement has to be generated between control and signal qubits. 

Quantum network mainly uses single-photon pulses as they represent practical realization of flying qubits, which can be used for long distance communication purposes. Several schemes for quantum router have been proposed, however, most of them do not satisfy all of the above six criteria. In some of the experiments, control bit only takes classical states \unskip~\cite{122770:4130815,122770:4133069}, hence resulting in a  semi-quantum router. Many of the experiments deal with light-matter interaction \unskip~\cite{122770:4130321,122770:4130322,122770:4130925}, which is challenging for experimental realization. Chang \textit{et al.}\unskip~\cite{122770:4131117} have come up with the idea of entanglement based quantum router, which does not satisfy the condition (2), as in this case, the control information is quantum however, the signal information collapses after the routing process. Lemr \textit{et al.} have proposed methods for realizing quantum router, still they do not fulfill conditions (5) \unskip~\cite{122770:4131118} and (6) \unskip~\cite{122770:4129601}. For the first time, a scheme of a genuine quantum router has been proposed by Yuan \textit{et al.}\unskip~\cite{122770:4131120}, which holds all of the above conditions. Recently, researchers have shown various routing processes using different architectures \unskip~\cite{122770:4131828,122770:4131829,122770:4131830,122770:4131831,122770:4131832,122770:4131874,122770:4131875,122770:4131827}.  

IBM quantum experience now plays a significant role in quantum computing community for giving access to two five-qubit and one sixteen-qubit quantum processors named as ibmqx2, ibmqx4 and ibmqx5 respectively. A number of quantum tasks such as quantum algorithms \unskip~\cite{122770:4132094,122770:4132095,122770:4132096,122770:4132097,122770:4132098}, quantum error correction code \unskip~\cite{122770:4132100,122770:4132101,122770:4132948}, quantum state and gate teleportation \unskip~\cite{122770:4132102,122770:4132103}, quantum information theory \unskip~\cite{122770:4132104,122770:4132105,122770:4132853,122770:4132854,122770:4132855,122770:4132947,122770:4132852}, quantum simulation \unskip~\cite{122770:4132856,122770:4132857}, quantum machine learning \unskip~\cite{122770:4132858}, quantum artificial intelligence \unskip~\cite{122770:4132859}, quantum communication devices \unskip~\cite{122770:4132861,122770:4132945,122770:4132946}, have been realized using both five-qubit and sixteen-qubit quantum processors. Here, we consider three qubits, among which one is control and the other two act as the signal qubits. Control qubit stores the control information while controlling the routing of signal information stored in the signal qubits. After the routing process, the entanglement is generated between the control qubit and the other signal qubits, and the signal information is found to be well preserved. The routing process satisfies all the necessary conditions to be called as quantum routing and the quantum circuit represents as a quantum router for the superconducting qubits.  

\section{Results}

\subsection{The Scheme of Quantum Routing}
\begin{figure}
\centering
\includegraphics[scale=0.5]{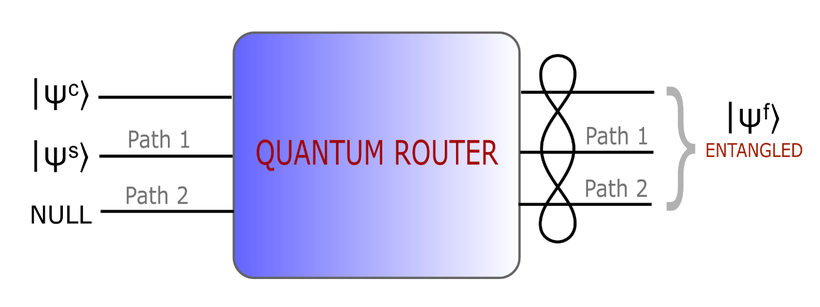} 
\caption{A schematic diagram illustrating the principle of a quantum router. The first qubit represents the control qubit which stores the control information ($|\Psi^{c}\rangle$), that directs the path of the signal information ($|\Psi^{s}\rangle$) initially stored in the second qubit. Here, path-2 contains no information, called as `Null' state. After the routing process, the information is found in a coherent superposition of two paths (the two signal qubits), i.e., path-1 and path-2. The final state, $|\Psi^{f}\rangle$ shows the generation of entanglement between the control qubit and the two paths.}
\label{Fig1}
\end{figure}

A schematic diagram for quantum router is depicted in Fig. \ref{Fig1}. The control qubit encodes the quantum information, $|\Psi_{c}\rangle=a|0\rangle+b|1\rangle$, with arbitrary coefficients a and b, where $|a|^2+|b|^2=1$. Two signal qubits are taken to store signal information whereas control qubit stores control information that will direct the path of the signal. The two signal qubits represent two possible paths for sending or storing signal information. Initially signal qubit-1 stores quantum data encoded as, $|\Psi_{s}\rangle_{1}=c|0\rangle+d|1\rangle$ with arbitrary coefficients c and d such that $|c|^2+|d|^2=1$. Here, the path-1 stores the signal information, and we encode $|+\rangle$ state in the signal qubit-2 to define `Null' state, that means there is no information on the signal qubit-2. We are considering the case where signal information is stored in the path-1, however, path-2 does not contain any information about the signal. For a classical router, the control qubit is either in state $|0\rangle$ or $|1\rangle$, hence the signal information is found either in signal qubit-1 (path-1) or in signal qubit-2 (path-2) according to the state of control. However, in quantum router, as the control qubit is in superposition of both $|0\rangle$ and $|1\rangle$, it is natural to find the signal information in a coherent superposition of both the paths after the routing process. 

\begin{figure}
\centering
\includegraphics[scale=0.3]{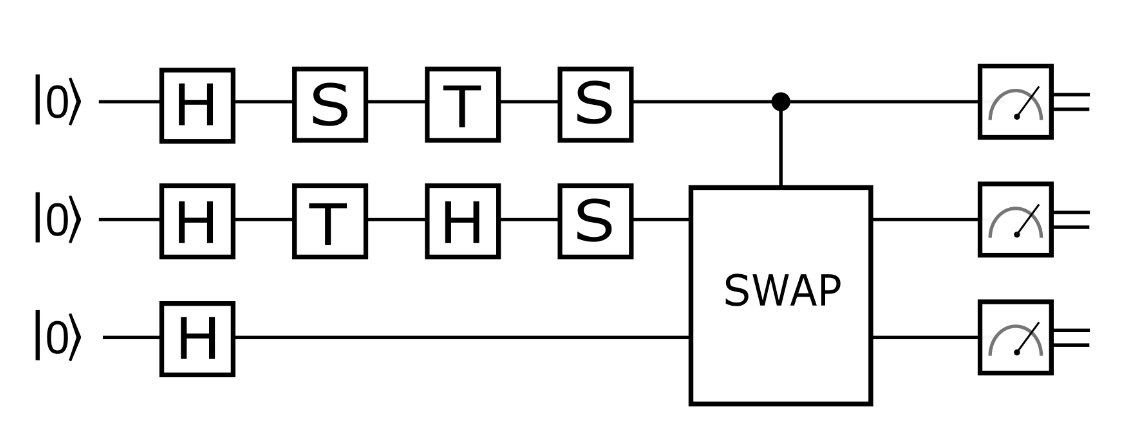} 
\caption{The first qubit represents the control qubit which is in a superposition state, $\frac{|0\rangle-e^{i\pi/4}|1\rangle}{\sqrt{2}}$. The signal information, $|\Psi_{s}\rangle=cos(\pi/8)|0\rangle+sin(\pi/8)|1\rangle$, is stored in the second qubit. The third qubit is in $|+\rangle$ state, which is conventionally taken to be a `Null' state. After the controlled-swap operation, the final state, $|\Psi_{f}\rangle=\frac{1}{\sqrt{2}}(|0\rangle|\Psi_{s}\rangle|+\rangle-e^{i\pi/4}|1\rangle|+\rangle|\Psi_{s}\rangle)$, is found in a three qubit entangled state generating entanglement between the control and the two signal paths. It can be easily seen that the signal information, $\Psi_{s}$, is preserved after the routing process.}
\label{Fig2}
\end{figure}

\subsection{Derivation of quantum routing process from the circuit given in Figs. \ref{Fig2}}

\begin{figure}
\centering
\includegraphics[scale=0.3]{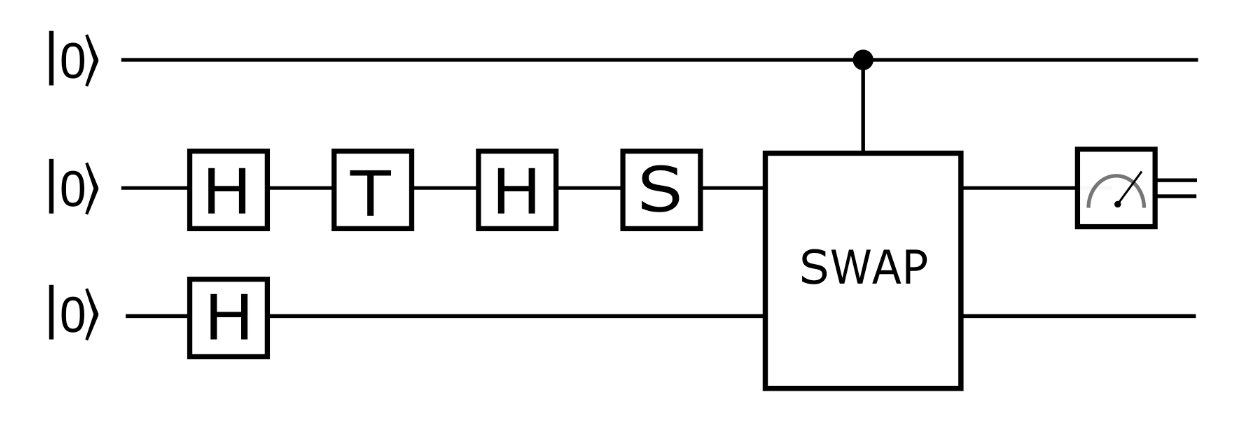} 
\caption{The first qubit, i.e., the control qubit is initially stored in $|0\rangle$ state. The signal information, $|\Psi_{s}\rangle=cos(\pi/8)|0\rangle+sin(\pi/8)|1\rangle$, is stored in the second qubit, called signal path-1 qubit. The third qubit is in $|+\rangle$ state (called signal's path-2 qubit), which is conventionally taken to be a Null state. After the controlled-swap operation, the final state becomes $|\Psi_{f}\rangle=|0\rangle|\Psi_{s}\rangle|+\rangle$, which implies the signal information is stored in the second qubit, preserving the signal information.}
\label{Fig3}
\end{figure}

\begin{figure}
\centering
\includegraphics[scale=0.3]{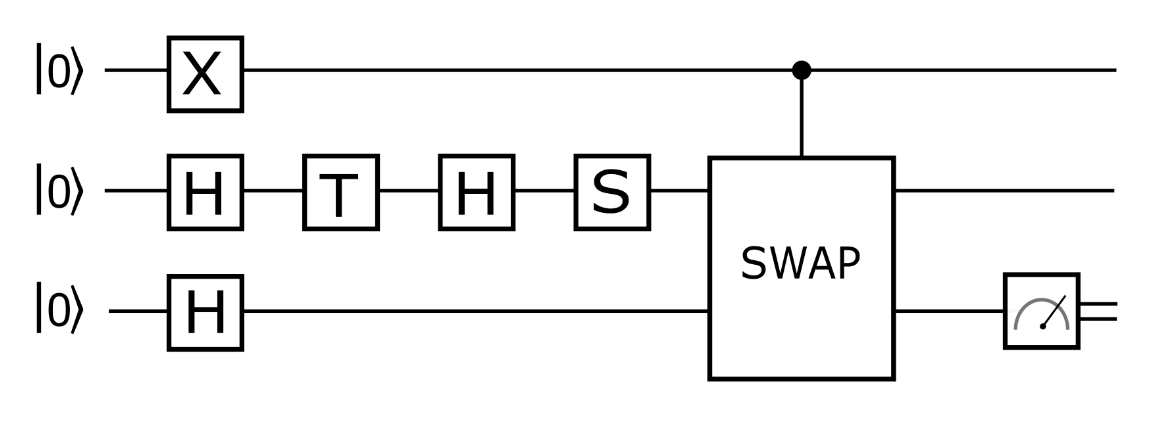} 
\caption{The first qubit, i.e., the control qubit is initially stored in $|1\rangle$ state. The signal information, $|\Psi_{s}\rangle=cos(\pi/8)|0\rangle+sin(\pi/8)|1\rangle$, is stored in the second qubit, called signal path-1 qubit. The third qubit is in $|+\rangle$ state (called signal path-2 qubit), which is conventionally taken to be a Null state. After the controlled-swap operation, the final state becomes $|\Psi_{f}\rangle=|1\rangle|+\rangle\Psi_{s}\rangle$, which implies the signal information is stored in the third qubit, preserving the signal information after being controlled by the control qubit.}
\label{Fig4}
\end{figure}

In the Fig. \ref{Fig2}, after applying H, S, T and S gates on the state $|0\rangle$, the control information becomes, $|\Psi_{c}\rangle=\frac{|0\rangle-e^{i\pi/4}|1\rangle}{\sqrt{2}}$. Similarly, the signal information can be calculated as, $|\Psi_{s}\rangle=cos(\pi/8)|0\rangle+sin(\pi/8)|1\rangle$. The initial state of the whole system can be written as, $|\Psi\rangle=|\Psi_{c}\rangle|\Psi_{s}\rangle|+\rangle$. 
After applying controlled-swap operation, the final state becomes, $|\Psi_{f}\rangle=\frac{1}{\sqrt{2}}(|0\rangle|\Psi_{s}\rangle|+\rangle-e^{i\pi/4}|1\rangle|+\rangle|\Psi_{s}\rangle)$. It is clearly seen that, the signal information is in path-1 when the control information is in state $|0\rangle$, and the signal information is in path-2, when the control information is in the state $|1\rangle$. It is also observed that the control qubit is entangled with the signal paths, i.e., it controls the path of the signal information. Such entanglement between control information and signal paths is the key mechanism to realize quantum transistor and quantum random access memory \cite{122770:4129685}. As in a classical router, the carried signal information remains preserved, hence a quantum router should follow this process, i.e., after the routing process, the signal information should be preserved. The preservation of the signal information has been shown by considering the following two cases (See Figs. \ref{Fig3} \& \ref{Fig4}), where the initial control information has been taken separately as both in $|0\rangle$ and $|1\rangle$ state. It is observed that, for case-1, i.e., the control information is in $|0\rangle$ state, the signal information is routed through the path-2, and in the case-2, i.e., the control information is in $|1\rangle$ state, the signal information is found in the path-2. It can be concluded from the above observation that the control information decides the route of the signal information. Hence, the whole scheme describes the working of a quantum router.            

\begin{figure}
\centering
\includegraphics[scale=0.5]{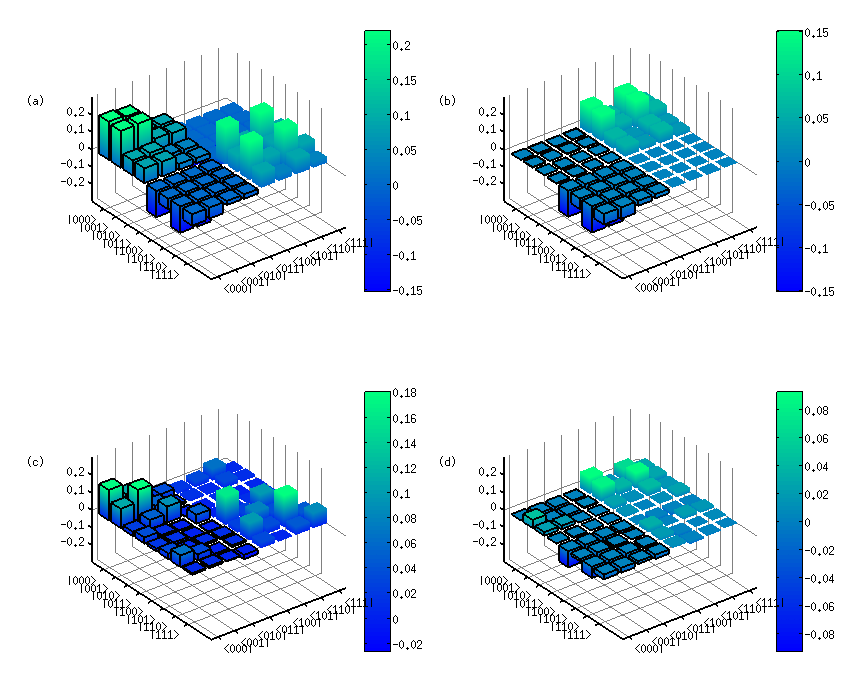} 
\caption{The figure illustrates the density matrices representing entanglement generation between the control qubit and the two signal paths. (a), (b) represent the real and imaginary part of the simulational density matrices, while (c), (d) representing the real and imaginary parts of reconstructed density matrices obtained from run results. The fidelity of the experimental results is found to be 0.9799.}
\label{Fig5}
\end{figure}

\begin{figure}
\centering
\includegraphics[scale=0.5]{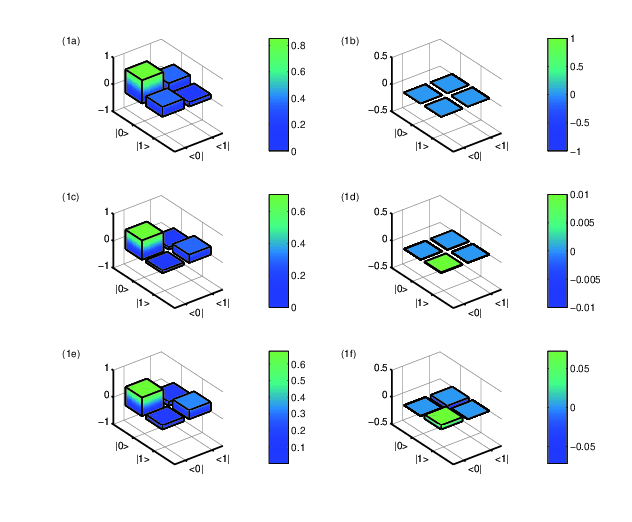} 
\caption{The density matrices represent the signal information. The theoretical and experimental results are compared. (a), (b) represent the real and imaginary part of the theoretical density matrices, while (c), (d) representing the real and imaginary parts of reconstructed density matrices obtained from run results, for the case when the control information is in $|0\rangle$ state. (e), (f) represent the real and imaginary part of the reconstructed density matrices for the case when the control qubit is in state $|1\rangle$.}
\label{Fig6}
\end{figure}

All the above quantum circuits are designed in IBM Quantum Experience interface using single-qubit gates (H, S, T, $T^{\dagger}$, X) and two-qubit quantum gate (CNOT gate). To verify the quantum nature of the quantum router, we confirm the coherence between the two signal paths and the generation of entanglement between the control qubit and signal paths provided they were initially in product states. We prepare the control qubit in $|\Psi_{c}\rangle$ state in a superposition state, by sequentially operating H, S, T and S gates on $|0\rangle$ state. Similarly, the signal information, $|\Psi_{s}\rangle$ is prepared by sequentially operating H, T, H and S gates on $|0\rangle$ state. Then it is stored in the signal path-1 qubit. A Hadamard gate is applied on the signal path-2 qubit showing the `Null' state, i.e., having no information about the signal. The output for the control qubit should be in an entangled state with the signal path for the ideal case. We verify the entanglement by performing measurement based three qubit state tomography process. We measure the state with 63 different measurement bases by taking 8192 number of shots. The density matrices for both the simulated and run result are plotted for comparison. Both the real and imaginary parts of reconstructed density matrices are shown in Fig. \ref{Fig5}. It is observed that the fidelity for the experimental density matrices are calculated to be 0.9799. Another important quantum nature of quantum router is that it should preserve the signal information carried by the signal qubit path-1. For verifying the preservation of quantum data or signal information, we have performed single-qubit quantum state tomography for both the cases when control qubit does not carry any superposition state. Rather it carries only $|0\rangle$ or $|1\rangle$ state, the case of a classical router. Hence, the measurement is performed on the second and third qubit according to the state of the control qubit. The comparison of density matrices for the theoretical and run results are shown in Fig. \ref{Fig6}. The fidelities for the above cases are found to be 0.9840 and 0.9777 respectively. Both the above experiments are repeated repeated for different superposition states of control qubit and for storing arbitrary information in the signal paths. In all the cases, we confirm the generation of entanglement between the control qubit and the signal paths, and the preservation of signal information, which are the two main operations of a quantum router to show the quantum nature of it. 

\section{Methods}

\begin{figure}
\centering
\includegraphics[scale=0.5]{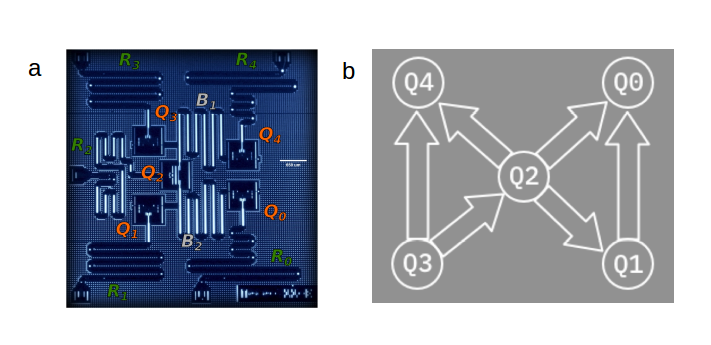}
\caption{(a) A schematic diagram of the chip layout of 5-qubit quantum processor `ibmqx4'. The chip is generally cooled in a dilution refrigerator at temperature 0.021 K. The connection of all 5 transmon qubits with the two coplanar waveguide (CPW) resonators are shown. With the resonance frequencies 6.6 GHz and 7.0 GHz, q[2], q[3], q[4] and q[0], q[1], q[2] are coupled with the two CPWs respectively. Individual qubits in the chip are controlled and readout by particular CPWs. (b) The CNOTs coupling map in the chip follows as, $\{q1 \rightarrow (q[0]), q2 \rightarrow (q[0], q[1], q[4]), q[3] \rightarrow (q[2], q[4])\}$, where $i \rightarrow (j)$ means $i$ and $j$ denote the control and the target qubit respectively for implementing CNOT gate. The errors in gates and readout are of the order $10^{-2}$ to $10^{-3}$.}
\label{Fig7}
\end{figure}

\begin{figure}
\centering
\includegraphics[scale=0.5]{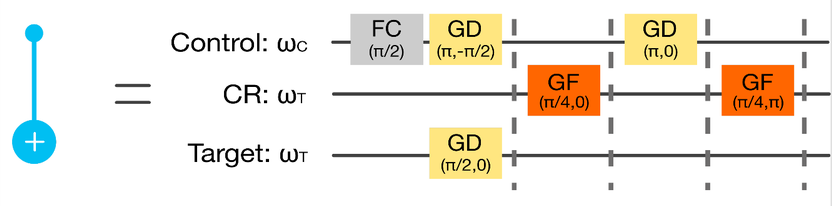} 
\caption{The figure illustrating the experimental construction of CNOT gate. Different frame change (FC), Gaussian derivative (GD) and Gaussian flattop (GF) pulses are applied with proper amplitude and angle parameters for implementation of CNOT gate.}
\label{Fig8}
\end{figure}

\textbf{Experimental Setup}: Some important experimental parameters of ibmqx4 chip are given in Table \ref{tab1}, where the readout resonator's resonance frequency, qubit frequency, anharmonicity, qubit-cavity coupling strength, relaxation time and coherence time are respectively denoted by $\omega^{R}_{i}$, $\omega_{i}$, $\delta_{i}$, $\chi$, $T_1$ and $T_2$. Fig. \ref{Fig7} (b) shows the connection and control of five superconducting qubits (q[0], q[1], q[2], q[3] and q[4]). The black and white lines represent the controls of the single-qubit and two-qubit controls provided by the coplanar wave guide (CPW) resonators. The qubits q[2], q[3], q[4] and q[0], q[1], q[2] are coupled via two superconducting CPWs, with 6.6 Hz and 7.0 Hz resonator frequencies respectively. Each qubit is controlled and read out by individual CPWs. The chip, 'ibmqx4' is stored in a dilution refrigerator at temperature around 0.021 K. The single-qubit gate error is of the order $10^{-3}$, and the multi-qubit and readout error is of the order $10^{-2}$. The gate errors are measured by using the process of randomized benchmarking. The experimental construction of CNOT gate is illustrated in Fig. \ref{Fig8}. 
 
\begin{table}
\centering
\begin{tabular}{ c c c c c c c }
\hline
\hline
Qubits & $\omega^{R^{\star}}_{i}/2\pi$ (GHz) & $\omega^{\dagger}_{i}/2\pi$ (GHz) & $\delta^{\ddagger}_{i}/2\pi$ (MHz) & $\chi^{\S}/2\pi$ (kHz) & $T^{||}_{1}$ ($\mu s$) & $T^{\perp}_{2}$ ($\mu s$)\\
\hline
q[0] & 6.52396 & 5.2461 & -330.1 & 410 & 35.2 & 38.1 \\
q[1] & 6.48078 & 5.3025 & -329.7 & 512 & 57.5 & 40.5 \\
q[2] & 6.43875 & 5.3025 & -329.7 & 408 & 36.6 & 54.8 \\ 
q[3] & 6.58036 & 5.4317 & -327.9 & 434 & 43.0 & 42.1 \\
q[4] & 6.52698 & 5.1824 & -332.5 & 458 & 49.5 & 19.2\\
\hline
\hline
\end{tabular}
$\star$ Resonance frequency, $\dagger$ Qubit frequency, $\ddagger$ Anharmonicity, $\S$ Qubit-cavity coupling strength, $||$ Relaxation time, $\perp$ Coherence time.
\caption{\textbf{The parameters of the device ibmqx4 are presented.}}
\label{tab1}
\end{table} 

\section{Conclusion}
To conclude we have demonstrated here the quantum nature of a quantum router by using a five-qubit quantum processor, `ibmqx4'. We have shown the working of quantum router by designing quantum circuit consisting of single-qubit and multi-qubit gates. We have clearly shown that the control information directs the signal information either to one path or superposition of several paths. We show the two main operations of a quantum router, i.e., entanglement generation between the control qubit and the signal paths, and preservation of signal information after the routing process. We have verified our experimental results by performing three-qubit and single-qubit quantum state tomography. We hope quantum router will find its significant application in areas like quantum network and quantum data processing. 

\section*{Acknowledgements}
BKB, TR and AG acknowledge the support of DST Inspire Fellowship. We are extremely grateful to IBM team for providing access to IBM Quantum Experience (QE). The discussions and opinions developed in this paper are only those of the authors and do not reflect the opinions of IBM or IBM QE team. 

\section*{References}

\bibliographystyle{iopart-num.bst}

\bibliography{\jobname}
 
\providecommand{\newblock}{}

\end{document}